\def\cf{{\it cf.\/}}

\documentstyle[prl,aps,twocolumn,epsf]{revtex}
\begin{document}
\draft
\title{kHz Quasi Periodic Oscillations in Low Mass X-ray Binaries \\  as 
Probes of General Relativity in the Strong Field
Regime} \author{Luigi Stella$^1$ and Mario Vietri$^2$}
\address{
$^1$ Osservatorio Astronomico di Roma, Via Frascati 33, 
00040 Monteporzio Catone (Roma) \\
E-mail: stella@heads.mporzio.astro.it ; 
affiliated to I.C.R.A \\
$^2$ Dipartimento di Fisica E. Amaldi, Universit\`a di Roma 3, 
Via della Vasca Navale 84, 00147 Roma 
\\ E-mail: vietri@corelli.fis.uniroma3.it}
\date{\today}
\maketitle
\begin{abstract}
We consider the interpretation of a pair of kHz Quasi Periodic Oscillations 
(QPOs) in the Fourier spectra of two Low Mass X-Ray Binaries, 
Sco X--1 and 4U1608-52, hosting an old accreting neutron star.  
The observed frequency difference of these QPOs decreaseas as their  
frequency increases, contrary to simple beat frequency models, which predict 
a constant frequency difference.  We show that the behaviour of these QPOs is 
instead well  matched in terms of the fundamental  frequencies
(in the radial and azimuthal directions) 
for test particle motion in the gravitational field of the 
neutron star,  for reasonable star masses, and nearly 
independent of the star spin. The radial 
frequency must be much smaller than the azimuthal one, testifying that
kHz QPOs are produced close to the innermost stable orbit. 
These results are not reproduced through the post--Newtonian (PN)
approximation of General Relativity (GR). kHz QPOs from
X-ray binaries likely provide an accurate laboratory for strong field GR.  
\end{abstract}

\narrowtext
\paragraph*{Introduction.}

The kHz QPOs in the X-ray flux of some 20 old accreting neutron stars in 
Low Mass X-Ray Binaries (LMXRBs) 
are the first involving timescales close to the
dynamical ones in the vicinity of a collapsed star \cite{klis}.
The frequency of these QPOs (from $\sim 0.33$ to $\sim 1.23$~kHz) displays 
changes which are often correlated with the instantaneous accretion rate.
A common phenomenon is the presence of a pair of 
kHz QPOs (with centroid frequencies $\nu_1$ and $\nu_2$)
drifting towards higher frequencies while keeping their difference 
$\Delta\nu \equiv \nu_2 - \nu_1$ approximately constant.  
The range of $\Delta\nu$ is also quite narrow across different sources 
($\Delta\nu \approx 250-360$~Hz).
During type I bursts from six of these 
sources, a nearly coherent signal at a frequency of $\approx 330 - 590$~Hz
was also detected. In a few cases the latter frequency is
consistent, to within the errors, with the frequency separation 
$\Delta\nu$, or twice its value $2\Delta\nu$. 

These results have been interpreted in terms of beat frequency
models,  BFMs (magnetospheric and sonic point) \cite{bfm} where
inhomogeneities at the inner disk boundary
are accreted at the  beat frequency between the local 
Keplerian frequency, $\nu_K$, and the  neutron star spin frequency,
$\nu_{spin}$. 
$\nu_K$ increases in response to increasing accretion rates 
and so does the QPO beat frequency $\nu_{b}=\nu_K-\nu_{spin}$.  
In these BFMs, the lower frequency kHz QPOs are  interpreted
as the beat frequency ($\nu_1 = \nu_b$) and the higher frequency kHz
QPOs as the Keplerian frequency ($\nu_2 = \nu_K$) at the inner disk
boundary. Therefore, $\Delta\nu$ yields the neutron star spin frequency 
($\nu_{spin} = \Delta\nu$). A signal at a frequency close to $\Delta\nu$ 
(or $2\Delta\nu$) is occasionally detected, but only during type I bursts. 
The inferred $\nu_{spin}$ spans a remarkably small range
($\sim 250-360$~Hz), far from the breakup limit of any neutron star
model. A direct consequence of the BFM interpretation, such a clustering of
spin frequencies across LMXRBs with very different average mass
transfer rates, orbital parameters, evolutionary histories
and magnetic field strenghts is surprising and calls for a separate
explanation  \cite{alternate}. 

Recent findings, however, undermine the applicability of simple BFMs 
to a growing number of LMXRBs with kHz QPOs. First, the separation 
$\Delta\nu$ decreases by 
$\sim80-100$~Hz as $\nu_2$ increases from $\sim800$ to $\sim1100$~Hz in 
Sco~X-1 \cite{klis1} and 4U1608-52 \cite{mendez}. 
Owing to poor statistics, a similar 
variation of $\Delta\nu$ in other sources would have remained
undetected \cite{psaltis}: the
behaviour of Sco~X-1 and 4U1608-52 might therefore be the rule.
Second, in 4U1636-536 
the frequency of the signal during bursts is  
inconsistent with  $\Delta\nu$ or its second harmonics \cite{mendez2}.

We propose here a substantially different 
scenario in which {\it both} kHz QPOs originate from the motion of matter
inhomogeneities
in the vicinity of the neutron star: as in BFMs, the higher frequency QPOs at 
$\nu_2$ are produced by the $\phi$-motion (i.e. the Keplerian motion) of 
inhomogeneities orbiting the inner disk boundary, while the 
lower frequency QPO signal at $\nu_1$ originates from the periastron precession 
at the inner edge of the accretion disk. Strong-field effects play a major 
role in determining the latter frequency. If this interpretation were 
confirmed, kHz QPOs would provide an unprecedented test for strong--field 
general relativity, GR.

\paragraph*{The model.}

We assume that the motion of the disk matter is dictated by the 
star's gravity alone.                                           
The gravitational field close to a neutron star is dominated by the 
Schwarzschild terms of the metric, with small corrections due
to the neutron star rotation.
For nearly circular orbits (eccentricity 
$e \simeq 0$) the coordinate frequency in the radial direction
$\nu_{S,r}$ as measured from an observer at infinity is given by 
\begin{equation}
$$\nu_{S,r} = \nu_{S,\phi} (1 - 6M/r)^{1/2}\;, $$
\end{equation}
where $r$ is the coordinate distance and $M$ the  mass of the compact object
(throughout this paper we use geometrical units $G=c=1$). 
The $\phi$-frequency retains its Keplerian value 
$\nu_{S,\phi} = \nu_K = (M/4\pi^2 r^3)^{1/2}$.

Eq. (1) shows that approaching the  marginally stable orbit ($r_{ms}=6M$),
$\nu_{S,r}$  vanishes, so that the precession frequency 
of the orbit's periastron 
$\nu_{per} \equiv \nu_\phi - \nu_r = \nu_{S,\phi} - \nu_{S,r} $ 
approaches the $\phi$-frequency. We propose here that, while the QPOs at   
$\nu_2$ arise from the $\phi$-frequency of matter orbiting close to the 
neutron star (as in BFMs), the QPOs at $\nu_1$ result 
from the strong-field periastron precession of the same matter. 
Therefore $\Delta\nu \equiv \nu_2-\nu_1 = \nu_{S,\phi} 
- (\nu_{S,\phi} - \nu_{S,r}) = \nu_{S,r}$. 
The curves in Fig.1 show $\nu_{S,r}$  as a function of $\nu_{S,\phi}$ for 
selected values of the neutron star mass, the 
only free parameter in the equations above. 
It is easy to see that $\nu_{S,r}$ decreases after a maximum of 
$\nu_{S,r}(max)= 358 (2M_{\odot}/M)$~Hz for a 
$\phi$-frequency of $\nu_{S,\phi} =  2\nu_{S,r}(max) 
\simeq  0.65\ \nu_{S,\phi}(r_{ms})$. 
The measured $\Delta\nu$ vs. $\nu_2$  for ten LMXRBs is also plotted.
It is apparent that for neutron star masses in the 2~M$_{\odot}$ range,
the simple model outlined above 
is in qualitative agreement with the measured 
values, including the  decrease of $\Delta\nu$ for increasing $\nu_2$ 
seen in Sco~X-1
and 4U1608-52.  

Developing a quantitative model involves allowing for finite orbital 
eccentricity and neutron star rotation. 
In order to calculate the $\phi$- and $r$-frequencies of bound orbits in the 
Schwarzschild metric, we define $\mu \equiv
M/a(1-e^2)$, where $a$ is the elliptic orbit semi--major axis. 
The radial frequency  is given by \cite{chandra} 
\begin{equation}
\nu_{S,r} = \frac{2\pi\nu_K  f(e,\mu)^{-1}}
{(1-e^2)^{3/2} \left((2\mu-1)^2 - 4\mu^2
e^2\right)^{1/2}} \;,
\end{equation}
where 
\begin{equation}
f(e,\mu) \equiv \int_0^{2\pi} \frac{d\!x [1-2\mu(3+e\cos x)]^{-1/2}}
{(1+e\cos x)^{2}[1-2\mu(1+e\cos x)]} 
\end{equation}
and $\nu_K = (M/4\pi^2 a^3)^{1/2}$. 
The azimuthal frequency $\nu_{S,\phi}$ is defined by an identical formula,
except that the integration in $x$ extends to $x_c$, 
implicitly defined by 
\begin{equation}
2\pi \sqrt{1-6\mu + 2\mu e} = \int_0^{x_c} \frac{d\!\gamma}
{(1-k^2\sin^2\gamma)^{1/2}}\;,
\end{equation}
with $k \equiv  4\mu e/(1-6\mu+2\mu e)$. 

The effects of the neutron star rotation are twofold. The gravitomagnetic 
effects are estimated here
at the first order of the angular momentum in the Kerr metric, {\it i.e.}
the Lense-Thirring order. We have \cite{LT} 
$\nu_{LT,\phi}=-\nu_{LT,nod}/2$ and $\nu_{LT,r}=3 \nu_{LT,nod}/2$ , where
\begin{equation}
\nu_{LT,nod}=4\pi a' \nu_K^2 (1-e^2)^{-3/2}
\end{equation}
is the Lense-Thirring nodal precession frequency, $a'= 2\pi\nu_{spin} I/M$ the
specific angular momentum and $I$ the moment of inertia of the rotating star.

The neutron star rotation-induced oblateness gives rise to a quadrupole 
component of the gravitational field. For orbits close to the equatorial 
plane, the quadrupole corrections are 
$\nu_{Q,\phi}=-\nu_{Q,nod}(1-e^2)^{1/2}$ and 
$\nu_{Q,r}=\nu_{Q,nod}[2-(1-e^2)^{1/2}] $ where 
\begin{equation}
\nu_{Q,nod}=-3\nu_K\Phi_2(2Mr^2)^{-1} (1-e^2)^{-2}
\end{equation}
is the quadrupole nodal precession frequency and $\Phi_2$ the star's quadrupole
moment. Detailed numerical calculations and approximate formulae for
$\nu_{LT,nod}$ and $\nu_{Q,nod}$ 
are given in \cite{morsink} for a variety of equations of state, EsOS. 
The corresponding corrections to the $\phi$, $r$ and periastron frequencies are 
small (up to tens of Hz) also for fast spinning neutron stars. 
Note that nodal precession 
arises exclusively from the relevant 
Lense-Thirring and quadrupole terms.
 
The $\phi$- and $r$-frequencies obtained by summing the components discussed
above were fit to the $\nu_2$ and $\Delta\nu$ frequencies 
measured in Sco~X-1. Initially $M$ and $e$
were treated as parameters of the fit, while 
different frequencies were calculated within each model
by stepping $a$, with $\nu_{spin}=0$. 
These fits had $e\sim0$ and were similar to those in 
Fig.~1. Allowing for neutron star rotation (including the 
Lense-Thirring and quadrupole terms) gave improved fits only for 
very stiff EsOS, such as L (\cf \cite{morsink} and 
references therein), and large counterrotating spin frequencies. 
For example a model consisting of nearly circular orbits 
with $\nu_{spin}=-520$~Hz, and a $1.8$~M$_\odot$ neutron star with EOS L
gave a fit similar to that shown in Fig.~2. 
While viable, this model is not satisfactory, because, 
if the neutron stars in LMXRBs have been spun up by accretion torques, one 
expects the disk matter to rotate in the same direction as the neutron star. 
 
Since there is no a priori reason why matter in the innermost disk 
region should maintain a constant $e$ as $a$ varies, we stepped $e$ 
in order to obtain different frequencies. The neutron star equator, marginally 
stable orbit or magnetospheric radius (depending on which is largest) 
provide a natural inner boundary to the accretion disk and the periastron 
of its innermost orbits. With this geometry in mind we introduced the 
periastron distance $r_p = a (1-e)$ as the second parameter in the fit. The 
best fit obtained in this way for a non-rotating neutron star is shown in 
Fig.~2. The agreement with the measured frequencies is remarkable. The variation
of $e$ with  $\nu_{\phi}$ is also shown. The model has, for the two free
parameters, $M\simeq 1.90$~M$_{\odot}$ and $r_p \simeq 6.25$~$M$, 
a value close to the marginally stable orbit. 

When $\nu_{spin}$ is allowed a finite value, 
slightly improved fits are obtained. For acceptable values of the 
neutron star spin frequency, however, these would be 
hard to distiguish from the best fit in Fig.~2. For a (corotating)  
$\nu_{spin}$ of 300 and 600~Hz and EOS L, the best fit paramaters would be 
$M\simeq 1.94$ and 1.98~M$_{\odot}$, $r_p \simeq 6.18$ and $6.17$~$M$, 
respectively. 
This shows that the effects induced by the neutron star rotation through the 
Lense-Thirring and quadrupole terms are quite small, though non-negligible; 
they cannot be currently disentagled from the Schwarzschild terms 
without a serious risk of overinterpreting the data and model.
While consistent with it, our model does not provide 
{\it independent} evidence 
that Sco~X-1, and likewise other LMXRBs, contain a fast spinning neutron star. 

We have shown that the model we propose accurately 
reproduces the observations with a minimum number of free paramaters 
(two, $M$ and $r_p$) the values of which are well within the expected range.
The fits described above (plus a 
large number of others that we tried), together with analytical 
estimates, show that for no acceptable value of the neutron star parameters
(mass, spin, EOS) the Lense-Thirring and quadrupole corrections alone 
can produce a large enough decrease of $\nu_r$ over the relevant range of 
$\nu_{\phi}$. Also, one may wonder whether the accuracy of our fits 
reflects the properties of the
strong field Schwarzschild metric, or some lower order expansion 
would reproduce the observed frequencies comparably well. So we carried out 
the calculations as in \cite{chandra} with the approximate PN metric for a 
spherical star\cite{mtw}, and repeated the fits described above. The 
modelling of the Sco~X-1 data was very poor, the best fits lying outside the 
range of Fig.2. We conclude that the matching of the model and the data is to
be ascribed mainly to the strong--field Schwarzschild terms. 

A simple argument shows also that the strong field regime applies.
In the interpretation we propose, $\nu_r$, often referred to as the 
epicyclic frequency, is $\ll \nu_\phi$. 
This ordering is unusual, the epicyclic frequency in most cases 
being the larger. However, if one remembers that $4\pi^2\nu_r^2 \equiv
d^2V_{eff}/dr^2$, where $V_{eff}$ is the effective potential, one sees that
the above ordering is natural for orbits close to the marginally stable one,
for which the effective potential notoriously has an inflection point, 
$d^2V_{eff}/dr^2 = 0$. Thus the relevant orbits must be close to the innermost 
stable orbit.

\paragraph*{Discussion.}

The higher frequency kHz QPOs at $\nu_2$ are interpreted in terms of 
the orbital $\phi$-frequency of matter at the inner edge of the accretion 
disk both within BFMs and the model presented here. 
The similarity in the highest value of $\nu_2$ ($\sim 1.1-1.2\; kHz$)
across different sources and, in a few cases, the saturation of $\nu_2$ at
its highest value despite an increasing source flux have been interpreted by
several authors as signs that the inner disk edge reaches the marginally
stable orbit \cite{rms}.

If, as predicted in our model, the lower frequency kHz QPOs arise from 
the periastron precession of the same matter that produces the 
higher frequency kHz QPOs,
then the exciting perspective opens up of probing directly the 
predictions of GR in the vicinity of the marginally stable 
orbit.
Moreover, results which are 
difficult to reconcile with BFMs are readily interpreted. 
First, the decreasing $\Delta\nu$ for increasing $\nu_2$ observed in Sco X-1 
\cite{klis1} and 4U1608-52 \cite{mendez} 
(and compatible with measurements in all other 
kHz QPO LMXRBs) results naturally from the expected strong field 
behaviour of $\nu_r$ in the vicinity of the marginally stable orbit.  
Second, the fact that in all kHz QPO sources $\Delta\nu$ is around 
$250-360$~Hz is naturally interpreted in terms of the 
characteristic broad maximum of $\nu_r$ versus $\nu_\phi$ apparent in
Fig.~1 provided the neutron star is $\sim 2$~M$_{\odot}$.
The predicted $\nu_r$ and $\nu_{\phi}$ depend only weakly on $\nu_{spin}$. 
Therefore, the measurement of $\nu_{spin}$ in QPO LMXRBs
has to rely only on coherent pulsations (so far detected only in  
1SAX~1808.4-3658 \cite{wijnands}) or, perhaps,                    
the signals observed during 
type I bursts from (low luminosity) Atoll 
sources \cite{hasinger}. 
The range of neutron star spins ($\sim 330-590$~Hz)
inferred from $\nu_{burst}$ is wider and shifted towards higher values, 
compared to that obtained within BFMs from $\Delta\nu$. This 
helps reconciling the spin period distribution of LMXRBs with that of  
millisecond radio pulsars. 
Cyg~X-2 and GX~17+2, somewhat peculiar type I bursters, might soon 
provide the first spin measurements in (high luminosity) Z-sources.

The model described here requires a small eccentricity of the blobs' orbits 
at the inner disk edge in order to match in detail the Sco~X-1 
kHz QPOs, without invoking large counterrotating neutron star spins. 
Such an eccentricity might originate from a variety of effects such as the 
amplification of the blobs' radial oscillations due to a resonant interaction
with the rotating magnetic field of the neutron star\cite{VS} or quasi-elastic 
impacts of blobs with the neutron star magnetospheric boundary or the boundary 
layer at the neutron star surface. 
If eccentricity variations drive the observed changes of $\nu_1$ and $\nu_2$, 
then the correlation of $\nu_1$ and $\nu_2$ with the mass 
accretion rate inferred from observations requires an anticorrelation of the 
eccentricity and $\dot{M}$; the mechanism for this is presently unknown,
but one may speculate that in low--eccentricity encounters of the blobs
with the edge of the precipice there is more time for matter to fall over. 

A modulation at the periastron precession frequency can be produced through 
the usual mechanisms, namely orbit-modulated emission by self-luminous blobs or 
partially occulting 
blobs  at the innermost disk boundary (\cf also \cite{SV}). In the first case 
Doppler beaming and blueshift enhance emission by the blobs when
the periastron velocity 
(the highest along the orbit) is closest to the line of sight; 
in the second the fraction of 
radiation from close to the neutron star that is occulted is a maximum 
whenever occultation is produced by a blob at periastron, where
a blob subtends the largest solid angle relative to the neutron star. 
Both situations repeat themselves with frequency $\nu_\phi-\nu_r$, 
the periastron precession frequency. 
The harmonic content of the signal(s) observed at infinity might be 
affected by strong gravitational lensing in the vicinity of the neutron 
star\cite{bardeen}.

Finally we note that the present model has also 
implications for the recent suggestion that the 
$\sim 20-35$~Hz oscillations observed in several kHz QPO sources of the 
Atoll group, and perhaps also the $\sim 15-60$~Hz horizonthal branch 
oscillations of Z-sources, originate in the (Lense-Thirring dominated)
nodal precession of blobs lifted off 
the equatorial plane at the inner disk boundary\cite{SV}. 
By using the spin frequencies inferred from $\nu_{burst}$
(as opposed to $\Delta\nu$ in BFMs) the predicted 
nodal precession frequencies for Atoll sources are up to a factor of  
$\sim 2$ higher and match the observations more accurately\cite{morsink}. 
On the contrary, in the application of the nodal precession  
model to Z-type LMXRBs the (as yet unmeasured) spin frequency 
is to be regarded as a free parameter. 

In summary, the QPO phenomenon in LMXRBs is likely to reflect the 
fundamental frequencies of the relativistic motion of matter 
in the vicinity of the neutron star. As discussed in the present {\it Letter}, 
if the $\phi$-frequency and periastron precession 
frequency are directly observed as kHz QPOs, these  
yield an unprecedented diagnostic of GR in the strong-field regime.

\begin{figure}
\caption{ kHz QPO frequency difference $\Delta\nu$ versus higher
QPO frequency $\nu_2$ for ten LMXRBs 
(cf. $[4,5,6]$ and references therein). 
Error bars are not plotted for the sake of clarity.
The models give the $r$- and $\phi$-frequencies of matter in nearly circular 
orbit around a non-rotating neutron star, of mass  
2.2, 2.0 and 1.8 M$_{\odot}$.}
\label{fig1}
\end{figure}

\begin{figure}
\caption{ kHz QPO frequency difference $\Delta\nu$ versus higher 
QPO frequency $\nu_2$ in Sco X-1. 
The best fit model corresponds to the $r$- and $\phi$-frequencies 
of matter orbiting a non-rotating 1.90~M$_{\odot}$ neutron star
at a periastron distance of 6.25~$M$ (17.5 km). The line marked 
with $e$ gives the orbital eccentricity ($\times 1000$) 
as a function of the $\phi$-frequency. 
Note that $e$ approaches 0 for the highest observed $\nu_2$; however  
a similar fit would be obtained by forcing $r_p$ to a slightly   
smaller value such that $e$ remains finite, though small, for    
any observed $\nu_2$.    
The upper curve is obtained by 
using the same parameters as above for a neutron star 
spinning at 600~Hz and EOS L.}
\label{fig2}
\end{figure}

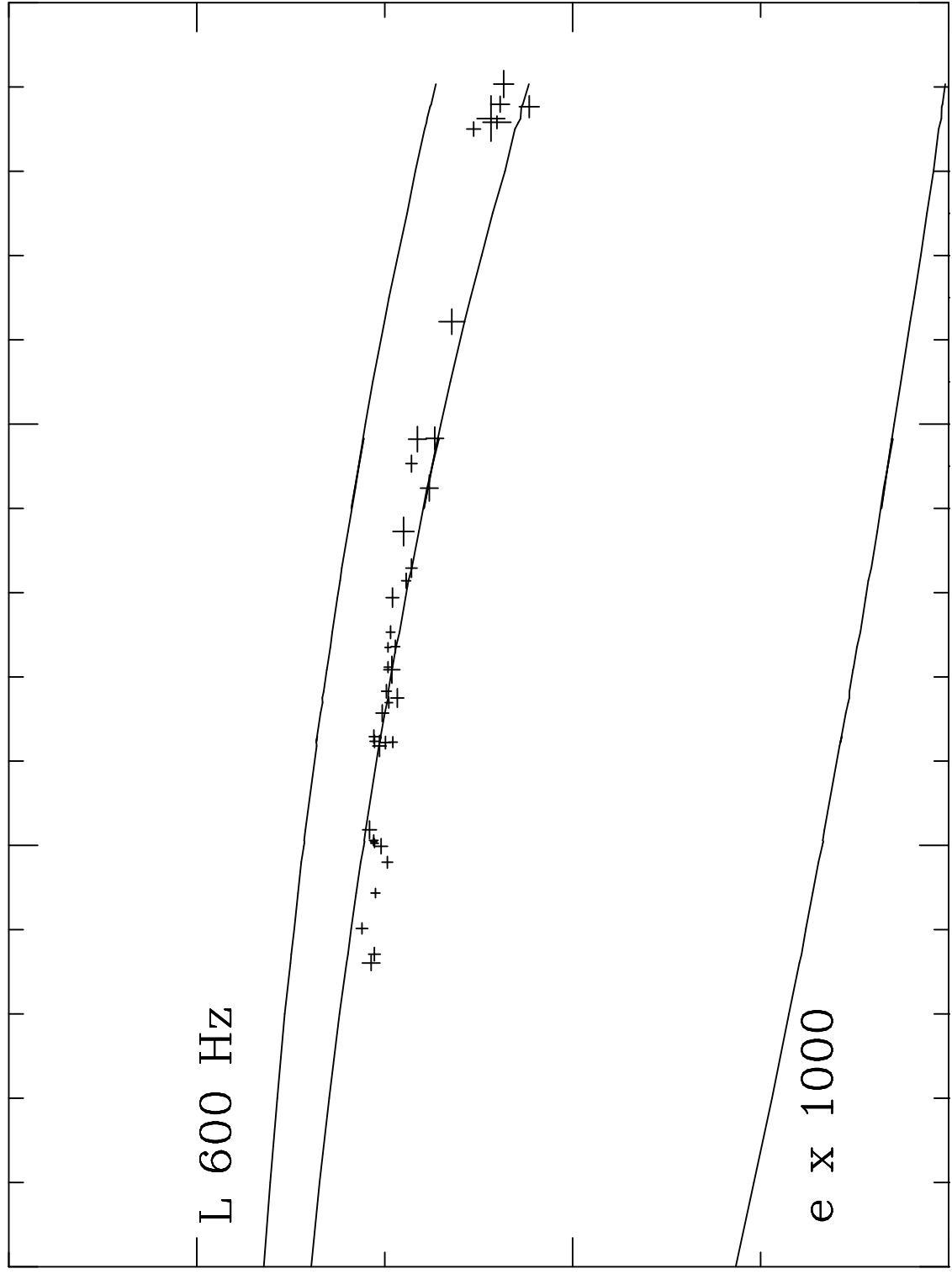
\end{document}